# Superconductivity in $Ru_{1-x}Sr_2GdCu_{2+x}O_{8-y}$ Compounds

P.W. Klamut*, B. Dabrowski, S. Kolesnik, M. Maxwell, J. Mais

Department of Physics, Northern Illinois University, DeKalb, IL 60115

**Abstract**

We report on the properties of new ruthenocuprates $Ru_{1-x}Sr_2GdCu_{2+x}O_{8-y}$ (x=0, 0.1, 0.2, 0.3, 0.4, 0.75) that extend the superconductivity found previously in $RuSr_2GdCu_2O_8$ ($T_c$=45 K) to the solid solution with varied Ru/Cu ratios. The compounds have been synthesized in high-pressure oxygen atmosphere. The maximum temperature of the superconducting transition is 72 K for the x=0.3 and 0.4 compositions. The reported behavior of magnetization at low temperatures can be qualitatively explained assuming a quasi-two-dimensional character of the superconducting regions in compounds studied.

------------

*corresponding author: email: klamut@niu.edu, fax: (815) 753-8565

**Introduction**

Recent reports of coexistence of superconductivity and ferromagnetism in ruthenocuprates $RuSr_2GdCu_2O_8$ (Ru-1212) [1, 2] and $RuSr_2(RE_{1-x}Ce_x)_2Cu_2O_{10}$, RE=Gd, Eu (Ru-1222) [3] have raised considerable interest in understanding the intrinsic properties of these layered materials. The crystal structure of ruthenocuprates can be described based on its similarity to $REBa_2Cu_3O_{7-y}$ (RE123) superconductors. For both Ru-1212 and -1222 the structure contains double $CuO_2$ planes separated by a single oxygen less Gd layer for Ru-1212, or a double fluorite $(RE_{1-x}Ce_x)_2$ block for Ru-1222. The Ru atoms, coordinating full octhaedra of oxygens, form the $RuO_2$ planes which replace the Cu-O chains present in RE123. The arrangement of Ru atoms resembles that in the $SrRuO_3$ itinerant ferromagnet ($T_c$=160 K) [4] or that in the $Sr_2RuO_4$ superconductor ($T_c$=1.5 K) for which the possible p-pairing of superconducting carriers was proposed [5]. The ferromagnetism with the transition temperature of 132 K that originates in the Ru sublattice was postulated for superconducting ($T_c \approx$30 K) Ru-1212 samples based on magnetization and muon-spin rotation experiments [2]. This observation raised the long standing issue of the conditions required for the coexistence of the two phenomena now in the class of high temperature superconducting compounds. Recent neutron diffraction experiments show that the dominant magnetic interactions present in $RuSr_2RECu_2O_8$ are of the antiferromagnetic (AFM) type with Ru moments forming the G-type antiferromagnetic structure [6, 7]. The ferromagnetism observed in these compounds was proposed to originate from the canting of Ru moments that give a net moment perpendicular to the c-axis [6, 7]. This description is similar to suggested for $Gd_2CuO_4$, a non-superconducting weak ferromagnet, where the distortions present in the CuO$_2$ plane permit the presence of the antisymmetric Dzialoshinski-Moriya superexchange interactions in the system of Cu magnetic moments [8]. The presence of both CuO$_2$ and RuO$_2$ planes in the structure of ruthenocuprates indicate the need for consideration of the model where the superconductivity is strictly constrained to the CuO$_2$ planes, whereas the magnetic properties originate in RuO$_2$ planes. Picket et al. [9] reported that strictly layered superconducting (SC) and ferromagnetic (FM) subsystems in Ru-1212 should be thin enough to allow 3D ordering by the coupling perpendicular to the layers but interacting weakly enough to permit superconductivity. To explain the absence of the apparent bulk Meissner state for Ru-1212 a model was proposed in which the superconducting order is modified into a fine structure in order to conform for the presence of ferromagnetic state [10]. Recently, we have reported that the partial substitution of trivalent Gd by Ce$^{4+}$ in RuSr$_2$Gd$_{1-x}$Ce$_x$Cu$_2$O$_8$ rapidly lowers $T_c$ and simultaneously raises $T_N$ to 145 K for the non-superconducting x=0.05 composition [11]. This decrease of $T_c$ is consistent with the doping effect observed for underdoped HTSC superconductors. The partial substitution of Nb$^{5+}$ into the Ru sublattice was found to lower both $T_c$ and $T_N$ for Ru-1212 [12].

Addressing the question how the superconducting properties of ruthenocuprates can be affected by the dilution of the magnetic sublattice of Ru, we attempted to partially substitute Ru with Cu ions. Previously, these kind of substitutions were extensively studied for several elements, Ga, Fe, Co, Ti, W, Mo and Re, and lead to the superconducting materials synthesized in air [13]. For Ru substitution we have found that the layered Ru-1212 type structure is stable only during synthesis at high pressure oxygen conditions. Here we report the primary properties of the new series of superconductors with the formula $Ru_{1-x}Sr_2GdCu_{2+x}O_{8-y}$. The series shows the systematic change of the superconducting and magnetic properties and should also promote a better understanding of the properties of the $RuSr_2GdCu_2O_8$ parent material. For the x=0.3 and 0.4 compositions the maximum critical temperatures were raised to 72 K. The reentrant behavior of the magnetization in the superconducting state at low temperatures suggests a quasi-two-dimensional character of superconductivity.

**Synthesis and Characterization**

Polycrystalline samples of $Ru_{1-x}Sr_2GdCu_{2+x}O_{8-y}$ (x=0, 0.1, 0.2, 0.3, 0.4, 0.75) were prepared by the solid-state reaction of stoichiometric $RuO_2$, $SrCO_3$, $Gd_2O_3$ and CuO. After calcination in air at 920°C the samples were ground, pressed into pellets, and annealed at 970°C in flowing oxygen. The samples were sintered at 1060°C for 10 hours in high pressure oxygen atmosphere (600 bar). The crystal structure was examined by the X-ray powder diffraction method using a Rigaku Inc. X-ray Diffractometer (CuKα radiation). The diffraction patterns showed that the Ru-1212-type structure formed for all compositions with traces of other impurity phases present (predominantly $SrRuO_3$). The samples were ground, pressed into pellets and annealed again in 600 bar of oxygen at 1085°C for 10 hours. Repeated annealing improved the phase purity of the material but did not change the temperatures of the superconducting transitions as verified by *ac* susceptibility measurements. The structure of all the samples was indexed in the tetragonal 4/mmm symmetry. The changes of the lattice parameters with x are presented in Fig.1. Both *a* and *c* decrease with the substitution of Cu for Ru in the Ru-O planes. This can indicate increased hole doping with x. The inset to figure 1 shows the x-ray diffraction pattern for the x=0.4 composition. An additional annealing of this sample in 200 bar of oxygen at 550°C did not change the temperature of the superconducting transition while annealing at 800°C in flowing air and in 1% of oxygen decreased $T_c^{on}$ from 72 to 55 and 43 K respectively. The effect of the oxygen content on the properties of x≠0 materials resembles qualitatively the properties of $YBa_2Cu_3O_{7-y}$ and is currently under investigation. The *ac* susceptibility, *dc* magnetization and resistivity data reported here were obtained for high pressure oxygen synthesized $Ru_{1-x}Sr_2GdCu_{2+x}O_{8-y}$ samples. The data were collected using a Quantum Design Physical Properties Measurement System.

**Results and Discussion**

Fig.2 presents the temperature dependencies of the field cooled (FC) and zero field cooled (ZFC) magnetization measured at approx. 15 Oe for the $Ru_{1-x}Sr_2GdCu_{2+x}O_{8-y}$ series. The compounds are supercondcuting at low temperatures, but FC magnetization at 4.2 K remains positive for all compositions except for x=0.75. The details of the superconducting transitions are discussed later in the text. The irreversibility of FC and ZFC branches observed below ~120 K and 100 K for x=0.1 and 0.2 samples, respectively (see insets in Fig.2) resemble the behavior of the magnetization observed below $T_N \approx 132$ K for $RuSr_2GdCu_2O_8$ [2] and thus should be attributed to the response of the Ru sublattice. However, the magnetization below these temperatures remains remarkably lower than for the x=0 parent material. The muon spin rotation experiments performed for the x=0.1 sample indicated that the increase of the relaxation rate observed below ~120 K should not be attributed to the bulk response of the material, contrary to what was observed for parent Ru-1212 [14]. It is not clear at present if the ZFC-FC irreversibility of magnetization curves found for x=0.1 and x=0.2 arises from compositional inhomogeneity (for example, the formation of Ru rich clusters in the Ru/Cu-O planes) or reflect the magnetic response of diluted $RuO_2$ planes. For x=0.3, 0.4 and 0.75 compositions we do not observe any irreversibility of the magnetization in the normal state. This indicates the absence of the long-range weak ferromagnetic order of the Ru sublattice above superconducting $T_c$. However, we should note that the AFM order, if is not accompanied by the FM component, would not be detected in this experiment.

Fig.3 shows an expanded view of *dc* magnetization and *ac* susceptibility at low temperatures. The unusual increase of the FC magnetization below the superconducting transition, which was already observed for $RuSr_2GdCu_2O_8$ [15], shows systematic behavior with increasing x. In Fig.3 we denote with $T_{c2}$ the onset temperature for the increase (x=0.3 and 0.4) and flat behavior (x=0.75) of FC magnetization below the temperature of its initial drop at $T_{c1}$. The $T_{c2}$ coincides with the temperature at which the *ac* susceptibility changes slope reflecting the increase of the bulk screening currents. Fig.4 compares the real parts of *ac* susceptibility measured for solid chunks and ground powder for the x=0.4 and x=0.75 samples. The diamagnetic screening of the bulk samples increases considerably below $T_{c2}$ indicating the onset of superconducting intergrain coupling and shows a complete shielding effect at 4.5 K. The much smaller diamagnetism measured for powder would usually indicate a small amount of the superconducting phase, or that the grain size (approx. 1 μm) is comparable to the penetration depth for this material. However, in the following discussion we present the arguments that the quasi-two-dimensional character of the superconducting regions can also account for this difference. By combining the *ac* susceptibility with the *dc* magnetization one can conclude that the increase of FC magnetization below $T_{c2}$ occurs when the shielding currents start to flow through the boundaries between superconducting regions.

The onsets of resistive transitions were found at $T_c^{on}$ = 45, 65, 70, 72, 72 and 62 K for x=0, 0.1, 0.2, 0.3, 0.4, and 0.75, respectively (see Fig.5). Fig.6 presents the superconducting resistive transitions for x=0.4, and x=0.75 samples measured in the magnetic fields of 0, 0.01, 0.05, and 0.1 T (solid lines) and 7 T (open circles). The resistivity develops a shoulder-like feature below $T_c$, the width of which remains very sensitive to small magnetic fields. With increasing x the width as well as the height of this shoulder decrease. If one correlates this behavior with smaller increase of FC magnetization below $T_{c2}$ for compositions with higher x (see Fig.3) it can be concluded that the interfaces between superconducting regions affect less efficiently the properties of compounds containing less Ru in the crystal structure. This pattern also seems to hold for the parent Ru-1212 for which the increase of FC magnetization is considerably more pronounced than for x≠0 compounds and its temperature $T_{c2}$ also correlates with the onset of bulk diamagnetism (compare with Ref. 15).

The temperature dependencies of ZFC *dc* susceptibility at 35, 45, 100, 200, 350, 500 and 1000 Oe for bulk x=0.75 sample are presented in Fig.7. The susceptibility decreases at the superconducting transition and then increases at lower temperatures for $H_{dc} \geq 50$ Oe. In spite of the zero resistance preserved in the high magnetic fields at low temperatures (see Fig.6 for ρ(T) at $H_{dc}$=7 T) the ZFC magnetization at 4.5 K remains negative only for magnetic fields lower than approx. 350 Oe. For the intermediate values of the magnetic field the additional decrease of the magnetization in the superconducting state is observed (denoted by dots in Fig.7). By comparing the FC and ZFC magnetization at 500 Oe measured for a chunk of the x=0.75 sample (Fig.8(a) – open squares) with the magnetization measured for powder (Fig.8(a).- closed squares) one can assign this decrease to the effect of enhanced diamagnetic response below the onset temperature for bulk superconducting screening. Interestingly, below this temperature the FC branch remains higher than the magnetization measured for powder. Similar behavior of FC magnetization was also observed for samples with smaller x at lower *dc* fields. Fig.8(b) shows the ZFC magnetization measured at $H_{dc}$=500 Oe for powder samples of x=0.4 and 0.75, and for the non-superconducting $GdBa_2Cu_3O_{7-y}$ (y≈0.8). By comparing these dependencies two contributions to the signal can be separated for x=0.4 and x=0.75: a diamagnetism related to superconductivity and the paramagnetic response of $Gd^{+3}$ ions. Paramagnetic behavior in the presence of superconductivity can be qualitatively understood assuming quasi-two-dimensional character of superconducting layers that are separated by non-superconducting regions. For polycrystalline samples with randomly oriented crystallines, the paramagnetic response would arise from the crystallines for which superconducting layers are oriented parallel to the external field that can penetrate the space between them. Similar effect was recently proposed to explain anisotropy of the magnetic susceptibility (for H ⊥ *ab* and H ∥ *ab*) observed for high oxygen deficient (i.e. strongly underdoped) superconducting $GdBa_2Cu_3O_{7-y}$ single crystals [16].

Fig 9. presents the M(H) dependencies for x=0.75 sample measured at 4.5, 20 and 50 K and magnetic fields changed between –500 Oe and 500 Oe. The first critical field at 4.5 K is estimated to be approx. 10 Oe (Fig.9(a)). The hysteresis loops can be interpreted as the superposition of the magnetic and superconducting components. The magnetic response at low temperatures arises from the paramagnetism of $Gd^{3+}$ ions. As this magnetic contribution decreases with increasing temperature, the magnetization remains negative for higher fields and presents complex hysteretic behavior (see Fig.9(b)). Above the temperature of the superconducting transition the M(H) dependence reflects only the paramagnetism of $Gd^{3+}$ ions (not shown). The detailed discussion of the low field magnetization behavior for $Ru_{1-x}Sr_2GdCu_{2+x}O_{8-y}$ will be presented separately.

The high magnetic field magnetization dependencies collected at 4.5 K for the whole series of $Ru_{1-x}Sr_2GdCu_{2+x}O_{8-y}$ are presented in Fig.10. In this experiment the magnetic field was changed between -6.5 and 6.5 T in 1000 Oe steps, so the data does not delineate the complicated M(T) behavior below 1000 Oe shown in Fig.9(a). The magnetizations for x≠0 samples are presented with solid lines that appear to converge to the curve obtained for non-superconducting $GdBa_2Cu_3O_{6.2}$ (open circles). The closed circles in Fig.10 represent the magnetization of the $RuSr_2GdCu_2O_8$ parent compound. By comparing parent and x≠0 samples it can be concluded that no additional contribution from the Ru sublattice to the measured signal is observed for diluted Ru sublattice, i.e. magnetic response is characteristic for the paramagnetic $Gd^{3+}$ ions as seen in $GdBa_2Cu_3O_{6.2}$. Larger magnetization values measured for $RuSr_2GdCu_2O_8$ indicates that only for this compound the long-range weak ferromagnetism of the Ru sublattice contributes to the high field magnetization by increasing its value by about 1 $\mu_B$. We should note that this contribution suggests almost complete ferromagnetic alignment of the Ru moments at high magnetic fields. However, the main magnetic contribution to the magnetization of RuSr$_2$GdCu$_2$O$_8$ arises from the paramagnetic system of Gd$^{3+}$ ions. This also can indicate the constrained dimensionality of the superconducting regions in this material.

In conclusion, we report that the new series of superconducting compounds, with the formula Ru$_{1-x}$Sr$_2$GdCu$_{2+x}$O$_{8-y}$, can be successfully synthesized at high pressure of oxygen. The maximum $T_c^{on}$=72 K (x=0.3 and 0.4) remarkably exceeds the superconducting transition temperature reported for RuSr$_2$GdCu$_2$O$_8$ ($T_c^{on}$=45 K). The signatures of the magnetic ordering of the Ru sublattice above the superconducting $T_c$ are present only for the x=0.1 and 0.2 samples. However, this feature can not be unambiguously attributed to the bulk of the material and detailed muon spin rotation and neutron diffraction experiments are necessary to resolve the magnetic behavior of Ru sublattice diluted with Cu ions. The observed reentrant behavior of magnetization below $T_c$ as well as its magnetic field dependence indicate that at low temperatures the magnetization becomes dominated by the paramagnetic response of the sublattice of Gd$^{3+}$ ions. This observation was qualitatively explained assuming a strong quasi-two-dimensional character of superconducting regions. Further studies of the nano-size characteristics of these compounds are required to investigate the suggested inhomogeneity of superconducting phase.

**Acknowledgement**

This work was supported by the ARPA/ONR and by the State of Illinois under HECA. It is a pleasure to acknowledge stimulating discussions with Drs. George Crabtree and Clyde Kimball.

**Literature**

[1]  L. Bauernfeind, W. Widder, H.F. Braun, Physica C **254**, 151(1995)

[2]  C. Bernhard, J.L. Tallon, Ch. Niedermayer, Th. Blasius, A. Golnik, E. Brücher, R.K. Kremer, D.R. Noakes, C.E. Stronach, E.J. Ansaldo, Phys. Rev. B **59** (1999) 14099

[3]  I. Felner, U. Asaf, Y. Levi, O. Millo, Phys. Rev.B **55**, R3374 (1997)

[4]  B.J. Kennedy, B. Hunter, Phys. Rev.B **58**, 653 (1998)

[5]  Y.Maeno, H. hashimoto, K. Yoshida, S. Nishizaki, T. Fujita, J.G. Bednorz, and F. Lichtenberg, Nature **372**, 532 (1994)

[6]  J.W. Lynn, B. Keimer, C. Ulrich, C. Bernhard, and J.L. Tallon, Phys. Rev. B **61**, R14964 (2000)

[7]  O.Chmaissem, J.D. Jorgensen, H. Shaked, P. Dollar, J.L. Tallon, Phys. Rev. B **61**, 6401 (2000); J.D. Jorgensen, O.Chmaissem, H. Shaked, S. Short, P.W. Klamut, B. Dabrowski, and J.L. Tallon, Phys. Rev.B., in press

[8]  J.D. Thompson, S.-W. Cheong, S.E. Brown, Z. Fisk, S.B. Oseroff, M. Tovar, D.C. Vier, and S. Shoultz, Phys. Rev. B **39**, 6660 (1989); P.W. Klamut, Phys. Rev. B **50**, 13009 (1994)

[9]  W.E. Pickett, R. Weht, and A.B. Shick, Phys. Rev. Lett. **83**, 3713 (1999)

[10] C.W. Chu, Y.Y. Xue, R.L. Meng, J. Cmaidalka, L.M. Dezaneti, Y.S. Wang, B. Lorenz and A.K. Heilman, cond-mat/9910056v3, Phys. Rev. Lett., in press

[11] P.W. Klamut, B. Dabrowski, J. Mais, M. Maxwell, Physica C **350**, 24 (2001)

[12] A. C. McLaughlin and J.P. Attfield, preprint

[13] B. Dabrowski, K. Rogacki, J.W. Koenitzer, K.R. Poeppelmeier, J.D. Jorgensen, Physica C **277**, 24 (1997); T. Den and T. Kobayashi, Physica C **196**, 141 (1992)

[14] P.W. Klamut, A. Shengelaya, R. Khasanov, I. Savic and H. Keller, B. Dabrowski, M. Maxwell, S. Kolesnik; unpublished, manuscript in preparation

[15] P.W. Klamut, B. Dabrowski, M. Maxwell, J. Mais, O. Chmaissem, R. Kruk, R. Kmiec, C.W. Kimball, Physica C **341-348**, 455 (2000)

[16] S. Kolesnik, T. Skoskiewicz, J. Igalson, M. Sawicki, and V.P. Dyakonov, Sol. State Comm. **97**, 957 (1996)

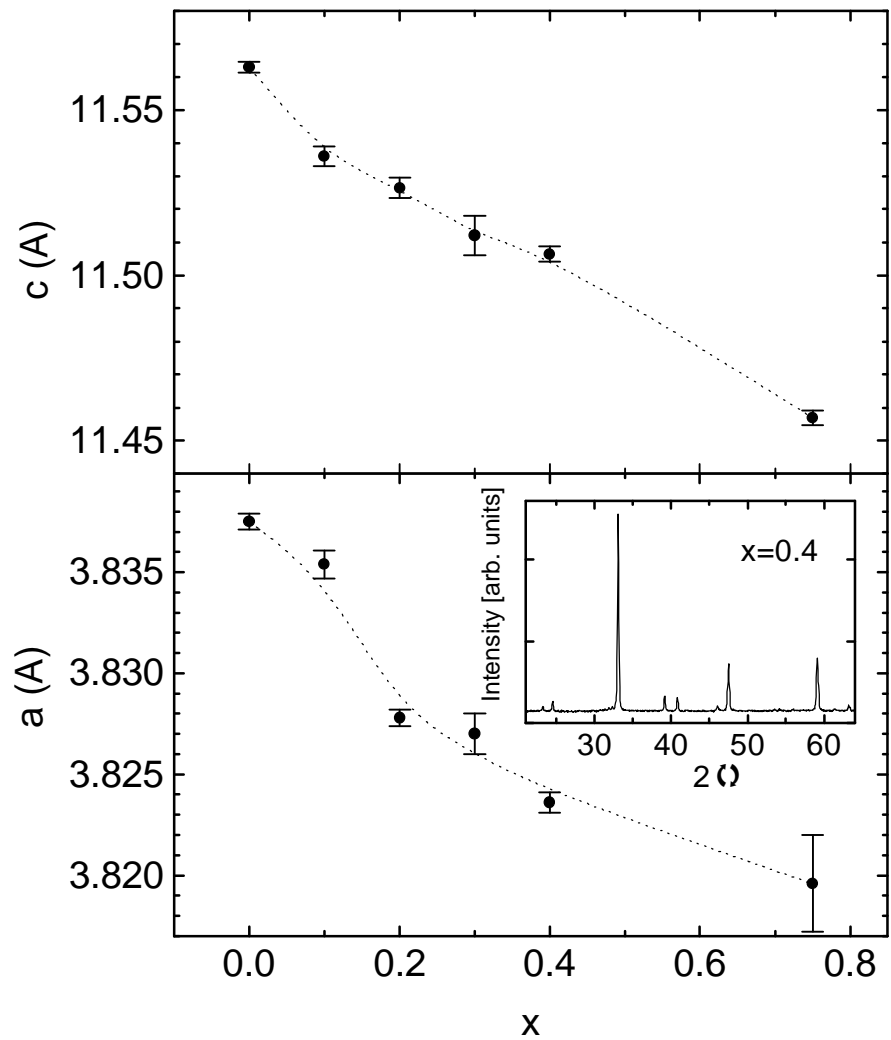

Fig.1. Lattice constants for the series of $Ru_{1-x}Sr_2GdCu_{2+x}O_{8-y}$ with x=0, 0.1, 0.2, 0.3, 0.4 and 0.75 (lines are guides to an eye). Inset presents the X-ray diffraction pattern for x=0.4 sample.

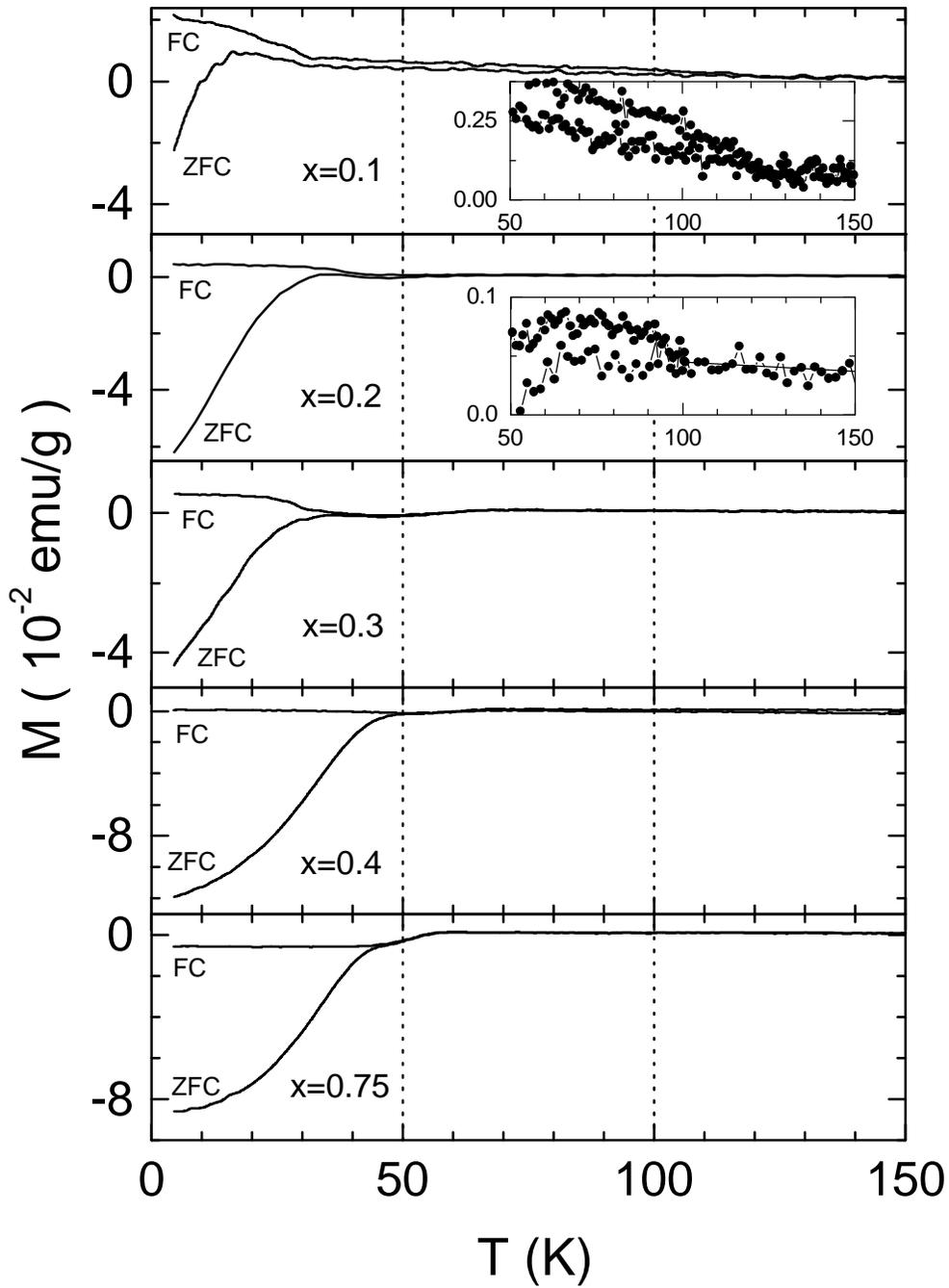

Fig.2. Temperature dependencies of the zero field cooled (ZFC) and field cooled (FC) *dc* magnetization ($H_{dc}$=approx. 15 Oe) for the $Ru_{1-x}Sr_2GdCu_{2+x}O_{8-y}$ series. For x=0.1 and 0.2 samples the corresponding insets show the onset of the irreversibility behavior in the normal state.

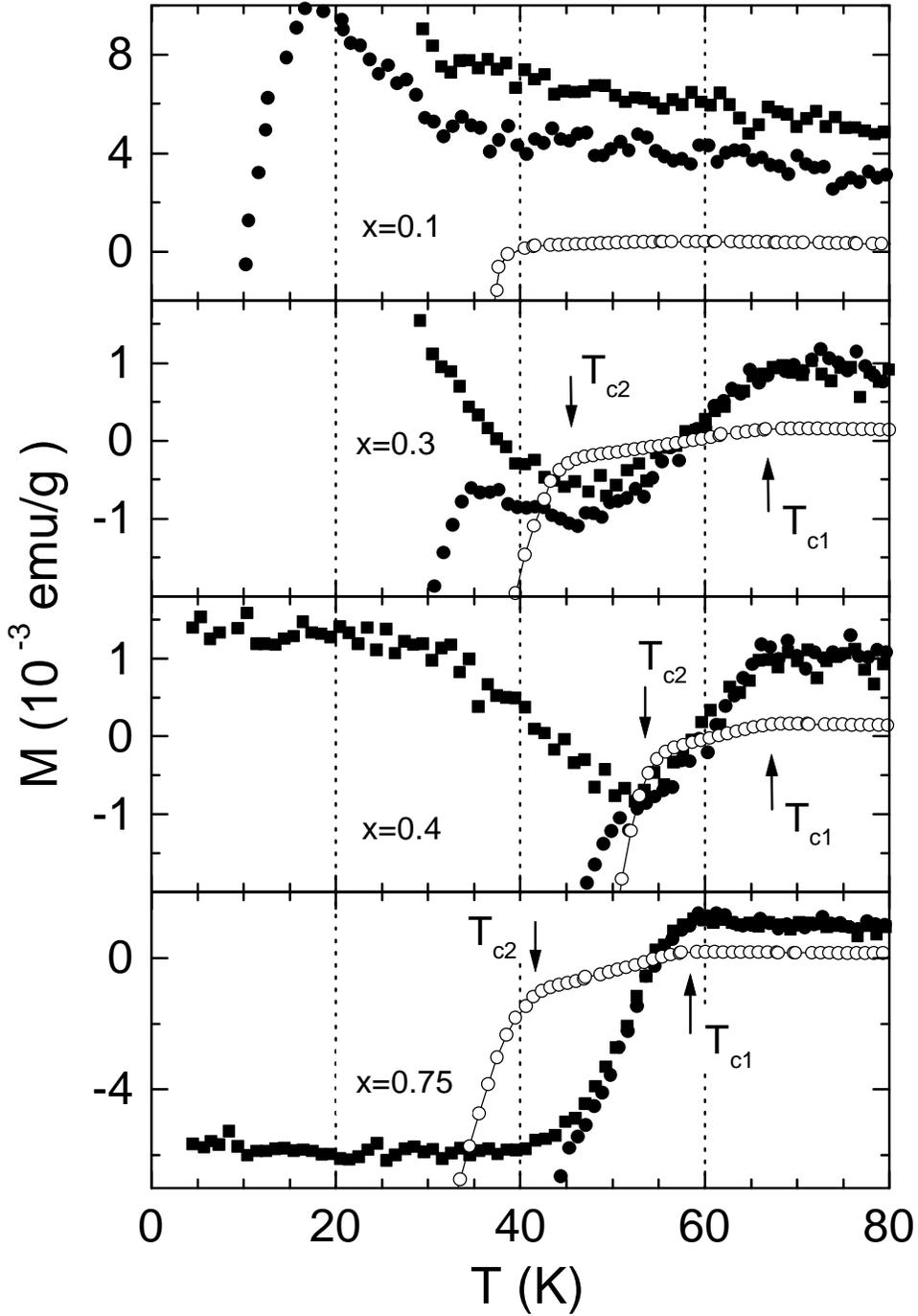

Fig.3. Temperature dependencies of the zero field cooled (closed circles) and field cooled (closed squares) *dc* magnetization ($H_{dc}$=approx. 15 Oe), and *ac* susceptibility (open squares, $H_{ac}$=1 Oe, 200 Hz) for the $Ru_{1-x}Sr_2GdCu_{2+x}O_{8-y}$ series. Expanded scale for the superconducting region, the spread of the data points within M(T) dependencies reflects the limitation of measurement's sensitivity. For the description of $T_{c2}$ and $T_{c1}$ see text.

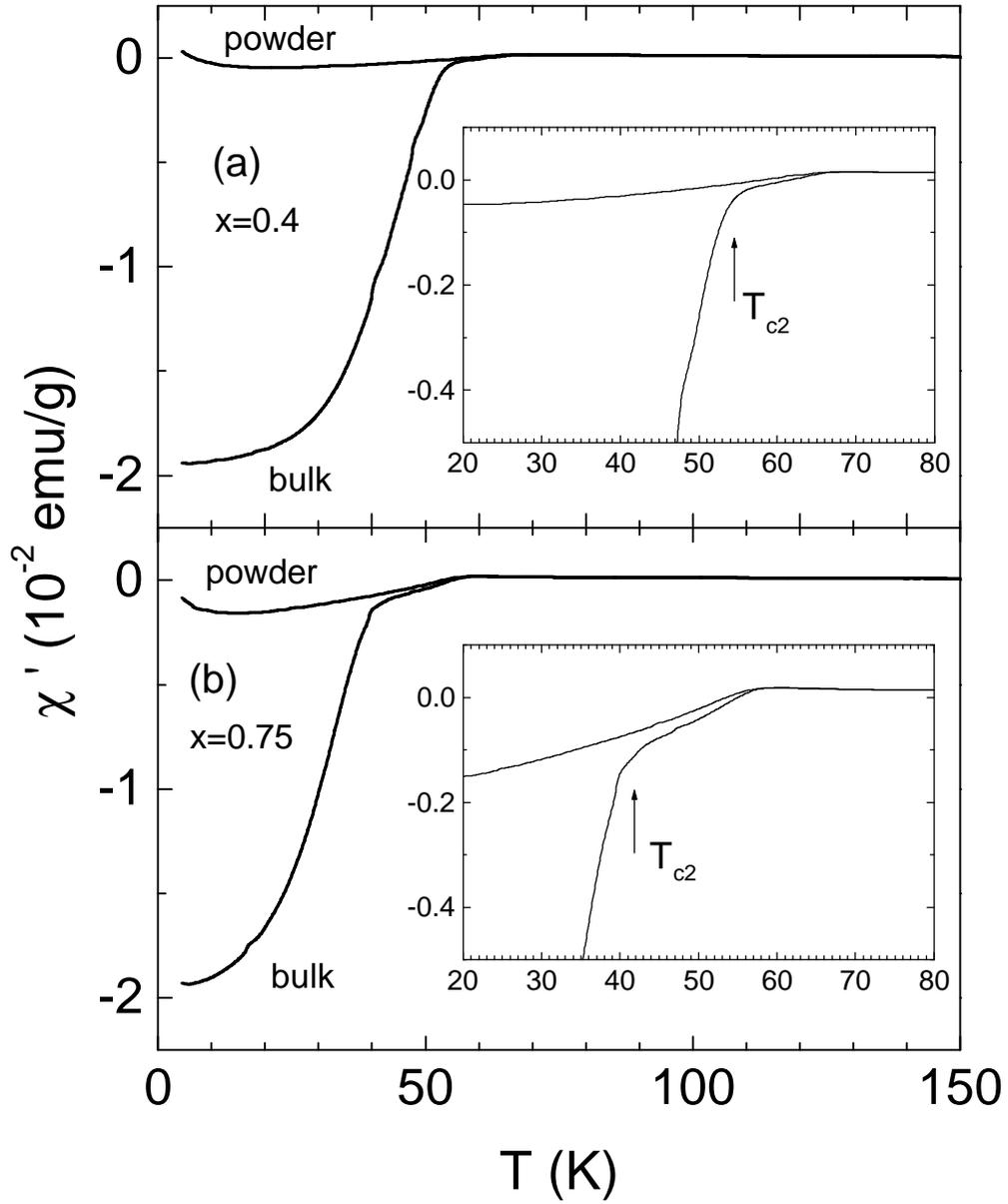

Fig.4. Temperature dependencies of the *ac* susceptibility for powder and bulk samples for x=0.4 (a) and x=0.75 (b). Insets present the onset of the transitions in the expanded scale. $H_{ac}$=1 Oe, f=200 Hz.

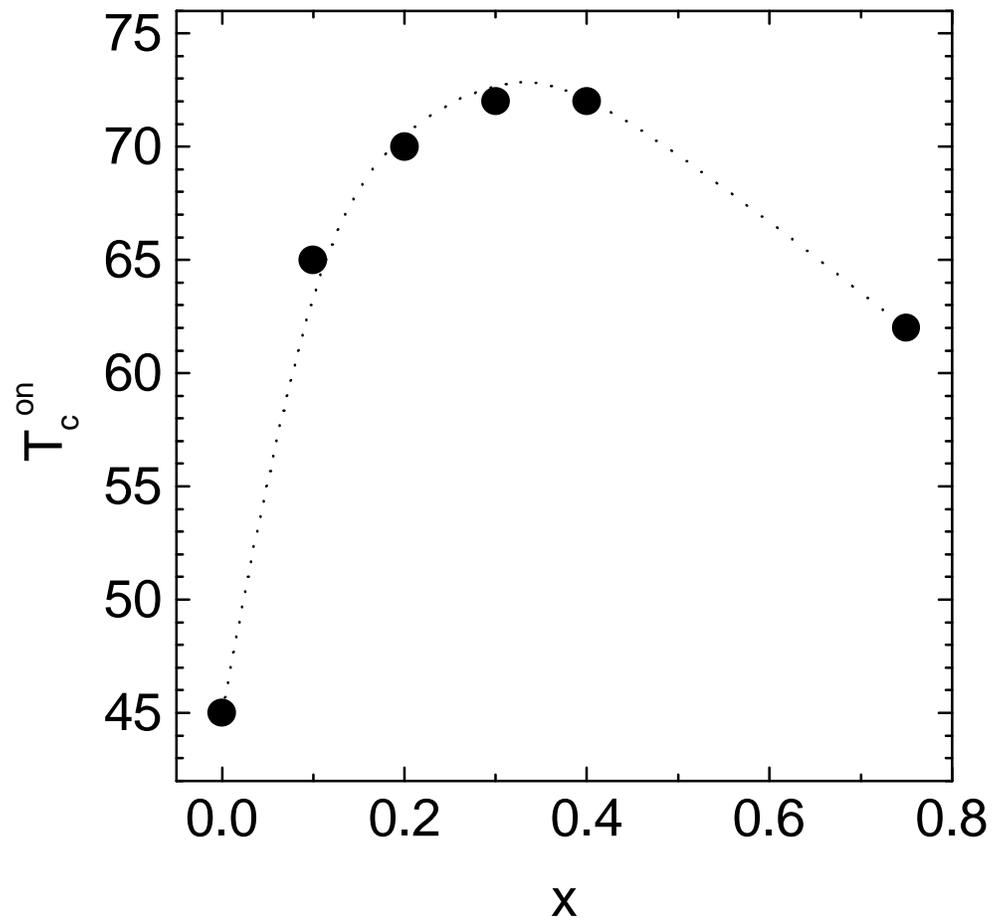

Fig.5. $T_c^{on}$ versus x for $Ru_{1-x}Sr_2GdCu_{2+x}O_{8-y}$ series.

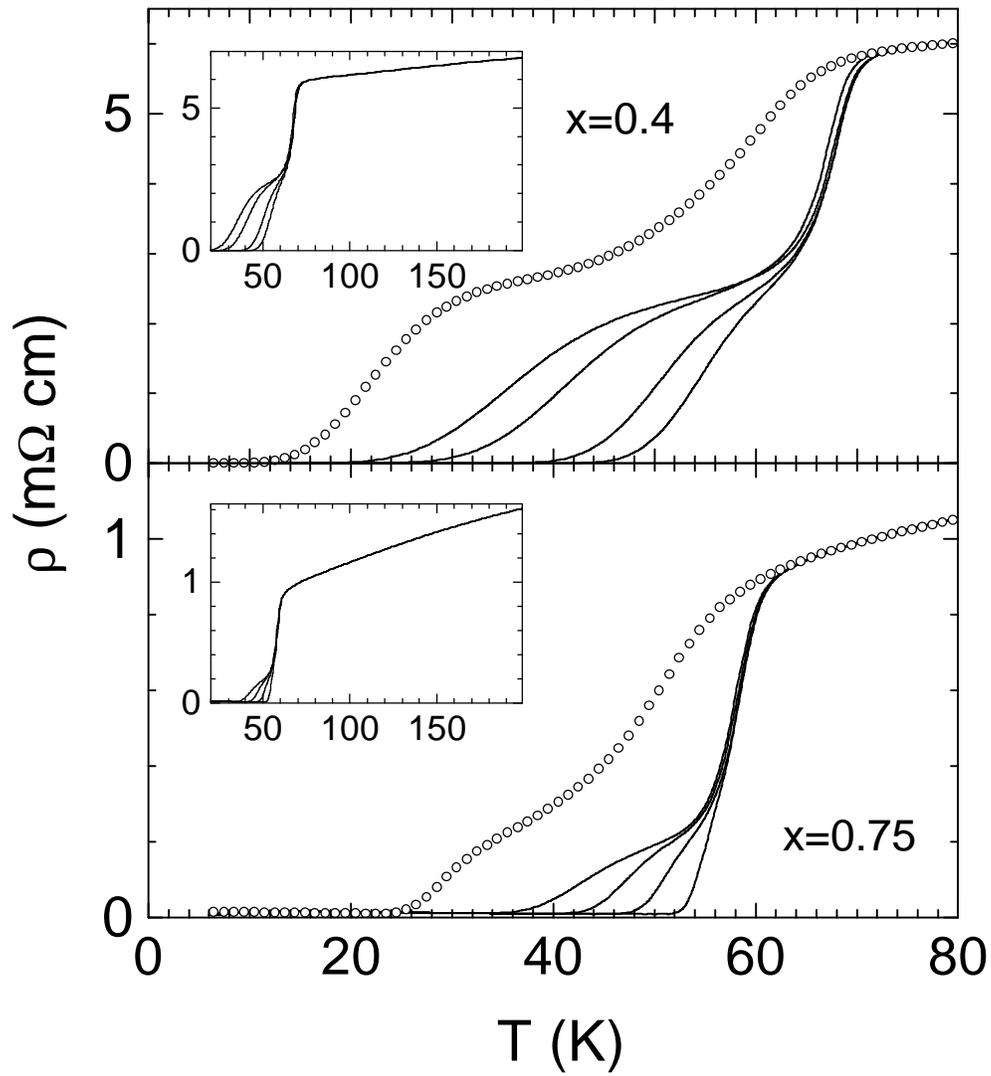

Fig.6. Resistivity transitions for x=0.4 and x=0.75 samples measured at 0, 100, 500, 1000 Oe (solid lines – the resistivity in the transition increases with field) and at 7 T (open circles). Insets show the normal state behavior. The data were collected on heating.

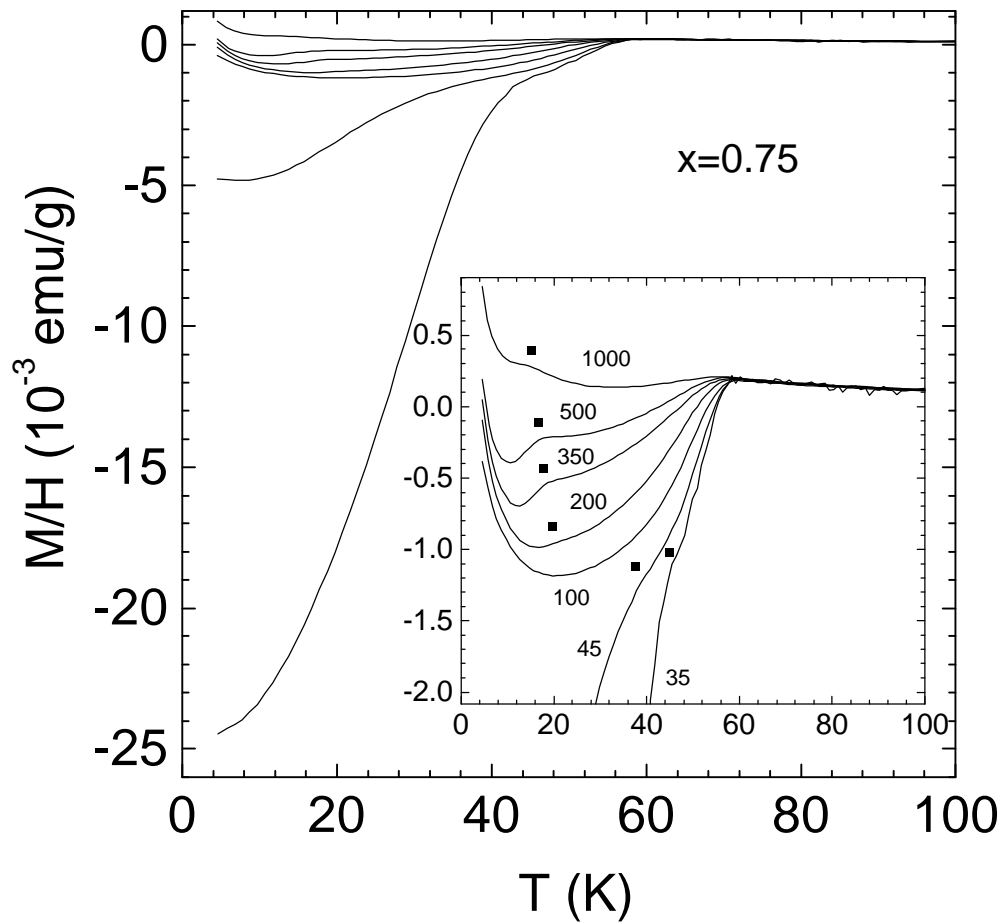

Fig.7. Temperature dependencies of ZFC *dc* susceptibility measured for x=0.75 sample at 35, 45, 100, 200, 350, 500 and 1000 Oe. Inset shows the behavior in the expanded scale. Closed squares mark temperatures for the onset of bulk superconducting screening (see text).

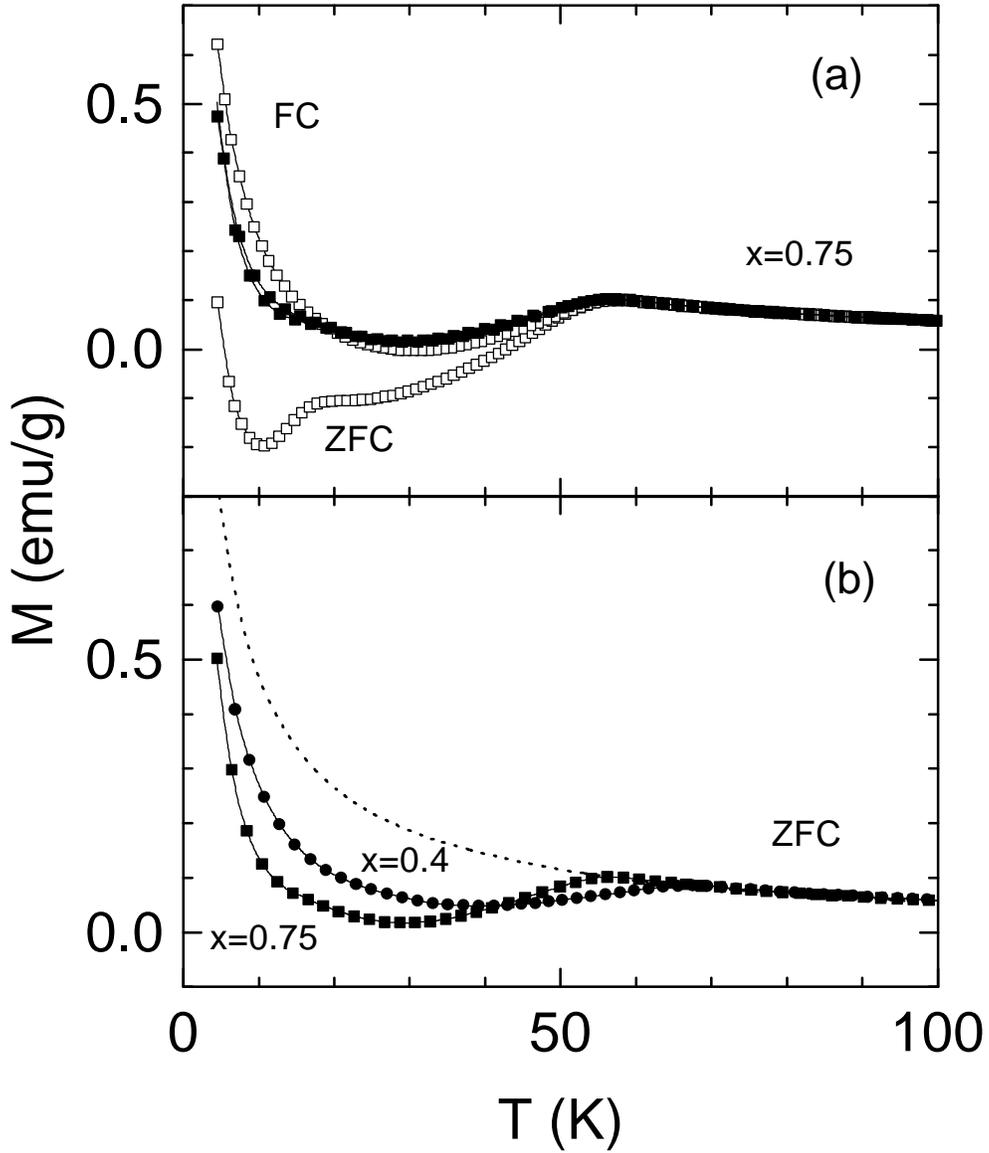

Fig.8. Temperature dependencies of ZFC and FC *dc* magnetization for powder (closed squares) and bulk (open squares) of x=0.75 sample (a); ZFC *dc* magnetization for x=0.4 (circles) and x=0.75 (squares) powder samples (b). Dotted line shows the corresponding dependence for non-superconducting $GdBa_2Cu_3O_{6.2}$ compound. $H_{dc}$=500 Oe.

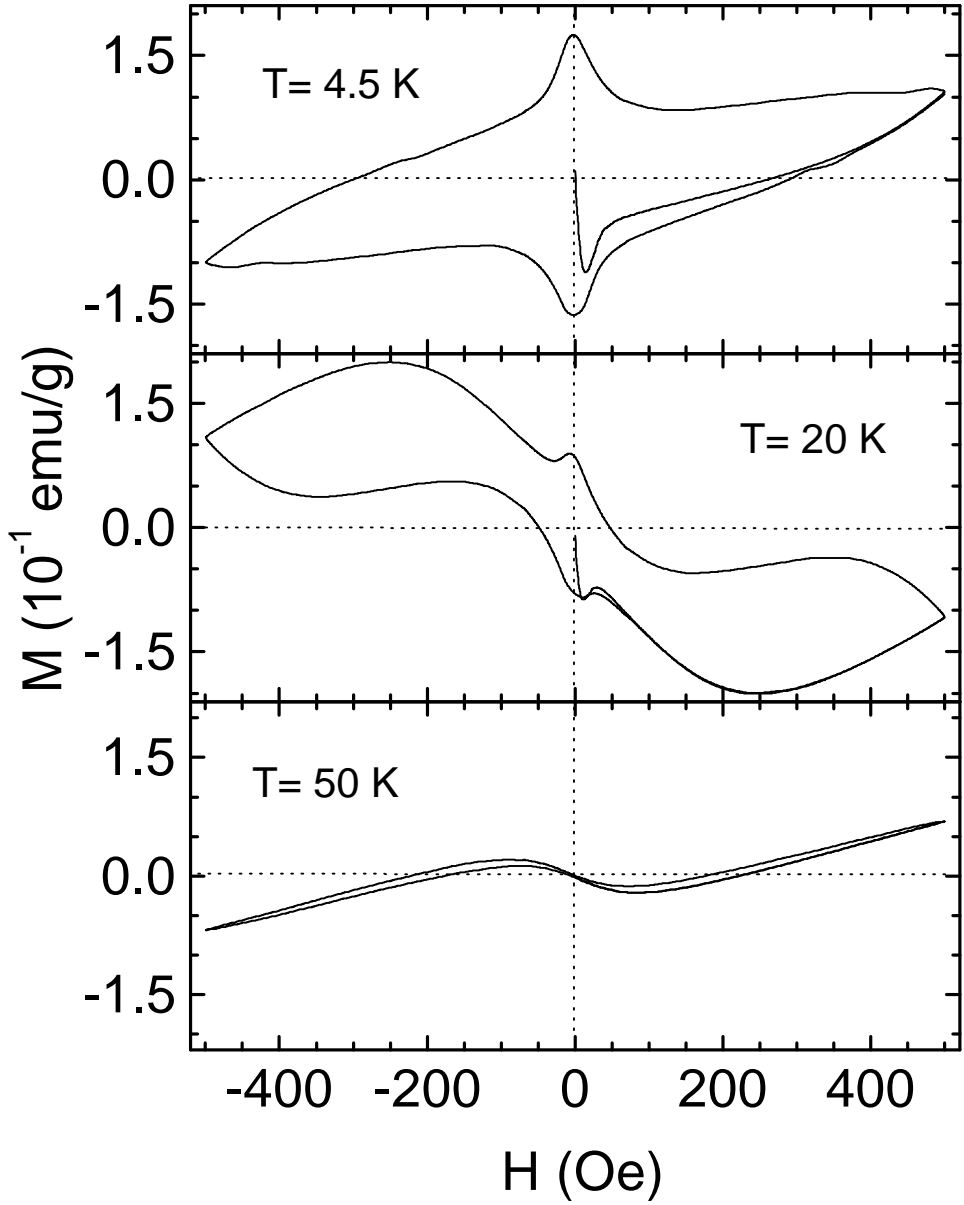

Fig.9. The magnetic field dependencies of the magnetization measured at (a) 4.5, (b) 20 and (c) 50 K for x=0.75 sample. The field was cycled between -500 and 500 Oe.

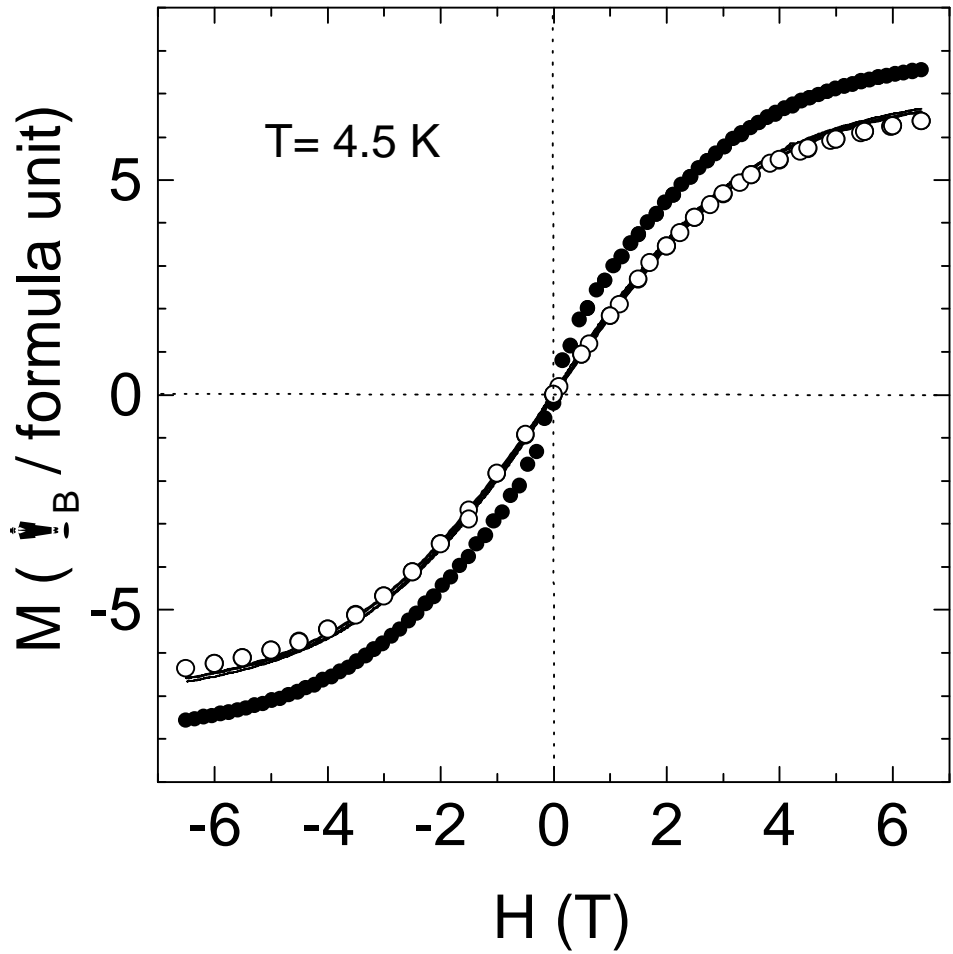

Fig.10. The magnetic field dependencies of the magnetization measured at 4.5 K for $Ru_{1-x}Sr_2GdCu_{2+x}O_{8-y}$ samples (solid lines converge to one curve). Closed circles show the behavior of superconducting $RuSr_2GdCu_2O_8$, open circles of non-superconducting $GdBa_2Cu_3O_{6.2}$.